\documentclass[12pt]{article}
\def\hybrid{\topmargin 0pt      \oddsidemargin 0pt
        \headheight 0pt \headsep 0pt
        \textwidth 17.5cm
        \textheight 25cm
        \voffset=-0.7cm
        \hoffset=-0.4cm
       \hoffset=-1.2cm
        \marginparwidth 0.0in
        \parskip 5pt plus 1pt   \jot = 1.5ex}
\catcode`\@=11
\def\marginnote#1{}

\newcount\hour
\newcount\minute
\newtoks\amorpm
\hour=\time\divide\hour by60
\minute=\time{\multiply\hour by60 \global\advance\minute by-\hour}
\edef\standardtime{{\ifnum\hour<12 \global\amorpm={am}%
        \else\global\amorpm={pm}\advance\hour by-12 \fi
        \ifnum\hour=0 \hour=12 \fi
        \number\hour:\ifnum\minute<10 0\fi\number\minute\the\amorpm}}
\edef\militarytime{\number\hour:\ifnum\minute<10 0\fi\number\minute}

\def\draftlabel#1{{\@bsphack\if@filesw {\let\thepage\relax
   \xdef\@gtempa{\write\@auxout{\string
      \newlabel{#1}{{\@currentlabel}{\thepage}}}}}\@gtempa
   \if@nobreak \ifvmode\nobreak\fi\fi\fi\@esphack}
        \gdef\@eqnlabel{#1}}
\def\@eqnlabel{}
\def\@vacuum{}
\def\draftmarginnote#1{\marginpar{\raggedright\scriptsize\tt#1}}

\def\draft{\oddsidemargin -0.1truein
        \def\@oddfoot{\sl preliminary draft \hfil
        \rm\thepage\hfil\sl\today\quad\militarytime}
        \let\@evenfoot\@oddfoot \overfullrule 3pt
        \let\label=\draftlabel
        \let\marginnote=\draftmarginnote
   \def\@eqnnum{{\rm (\theequation)}\rlap{\kern\marginparsep\tt\@eqnlabel}%
\global\let\@eqnlabel\@vacuum}  }

\newdimen\linethick  \linethick=0.4pt
\newdimen\hboxitspace    \hboxitspace=5pt
\newdimen\vboxitspace    \vboxitspace=5pt

\def\fr#1{%
\beq\new
\vcenter{
\hrule height\linethick
           \hbox{\vrule width\linethick
                 \kern\hboxitspace
                 \vbox{\kern\vboxitspace
                       \hbox{$\begin{array}{c}\displaystyle#1
          \end{array}$}%
                       \kern\vboxitspace}%
                 \kern\hboxitspace
                 \vrule width\linethick}%
           \hrule height\linethick}%
\eeq}

\newdimen\Squaresize \Squaresize=14pt
\newdimen\Thickness \Thickness=0.5pt

\def\Square#1{\hbox{\vrule width \Thickness
   \vbox to \Squaresize{\hrule height \Thickness\vss
      \hbox to \Squaresize{\hss#1\hss}
   \vss\hrule height\Thickness}
\unskip\vrule width \Thickness}
\kern-\Thickness}

\def\Vsquare#1{\vbox{\Square{$#1$}}\kern-\Thickness}

\def\numberbysection{\@addtoreset{equation}{section}
        \def\theequation{\thesection.\arabic{equation}}}
\numberbysection

\renewcommand{\theequation}{\thesection.\arabic{equation}}
\newcommand{\l@qq}[2]{\addvspace{2em}
 \hbox to\textwidth{\hspace{1em}\bf #1 \dotfill #2}}

\newcounter{app}

\def\app{\setcounter{equation}{0}
\def\theequation{\Alph{app}.\arabic{equation}}\par
   \addvspace{4ex}
   \@afterindentfalse
  \secdef\@app\@dapp}
\newcommand\@app{\@startsection {app}{1}{0ex}%
                                   {-3.5ex \@plus -1ex \@minus -.2ex}%
                                   {2.3ex \@plus.2ex}%
                                   {\normalfont\Large\bf}}

\def\@dapp#1{%
{\parindent \z@ \raggedright  \bf #1}\par\nobreak}
\def\l@app#1#2{\ifnum \c@tocdepth >\z@
    \addpenalty\@secpenalty
    \addvspace{1.0em \@plus\p@}%
    \setlength\@tempdima{2.5em}%
    \begingroup
      \parindent \z@ \rightskip \@pnumwidth
      \parfillskip -\@pnumwidth
      \leavevmode \bfseries
      \advance\leftskip\@tempdima
      \hskip -\leftskip
      #1\nobreak\hfil \nobreak\hb@xt@\@pnumwidth{\hss #2}\par
    \endgroup\fi}
\newcounter{sapp}[app]

\def\sapp{\def\theequation{\Alph{app}.\arabic{equation}}\par
   \@afterindentfalse
  \secdef\@sapp\@dsapp}
\newcommand\@sapp{\@startsection{sapp}{2}{\z@}%
                                     {-3.25ex\@plus -1ex \@minus -.2ex}%
                                     {1.5ex \@plus .2ex}%
                                     {\normalfont\large\bfseries}}

\def\@dsapp#1{%
{\parindent \z@ \raggedright  \bf #1}\par\nobreak}
\newcommand{\l@sapp}{\@dottedtocline{2}{1.5em}{3em}}

\def\titlepage{\@restonecolfalse\if@twocolumn\@restonecoltrue\onecolumn
     \else \newpage \fi \thispagestyle{empty}\c@page\z@
        \def\thefootnote{\fnsymbol{footnote}} }

\def\endtitlepage{\if@restonecol\twocolumn \else  \fi
        \def\th\tilde{\Lambda}(\Lambda)\tilde{\Lambda}(\Lambda)efootnote{\arabic{footnote}}
        \setcounter{footnote}{0}}  
\relax

\hybrid

\newtoks\@stequation

\def\subequations{\refstepcounter{equation}%
  \edef\@savedequation{\the\c@equation}%
  \@stequation=\expandafter{\theequation}
  \edef\@savedtheequation{\the\@stequation}
  \edef\oldtheequation{\theequation}%
  \setcounter{equation}{0}%
  \def\theequation{\oldtheequation\alph{equation}}}

\def\endsubequations{%
  \setcounter{equation}{\@savedequation}%
  \@stequation=\expandafter{\@savedtheequation}%
  \edef\theequation{\the\@stequation}%
  \global\@ignoretrue}

\makeatletter
\newdimen\normalarrayskip              
\newdimen\minarrayskip                 
\normalarrayskip\baselineskip
\minarrayskip\jot
\newif\ifold             \oldtrue            \def\new{\oldfalse}
\def\arraymode{\ifold\relax\else\displaystyle\fi} 
\def\eqnumphantom{\phantom{(\theequation)}}     
\def\@arrayskip{\ifold\baselineskip\z@\lineskip\z@
     \else
     \baselineskip\minarrayskip\lineskip1\baselineskip\fi}

\def\@arrayclassz{\ifcase \@lastchclass \@acolampacol \or
\@ampacol \or \or \or \@addamp \or
   \@acolampacol \or \@firstampfalse \@acol \fi
\edef\@preamble{\@preamble
  \ifcase \@chnum
     \hfil$\relax\arraymode\@sharp$\hfil
     \or $\relax\arraymode\@sharp$\hfil
     \or \hfil$\relax\arraymode\@sharp$\fi}}

\def\@array[#1]#2{\setbox\@arstrutbox=\hbox{\vrule
     height\arraystretch \ht\strutbox
     depth\arraystretch \dp\strutbox
     width\z@}\@mkpream{#2}\edef\@preamble{\halign \noexpand\@halignto
\bgroup \tabskip\z@ \@arstrut \@preamble \tabskip\z@ \cr}%
\let\@startpbox\@@startpbox \let\@endpbox\@@endpbox
  \if #1t\vtop \else \if#1b\vbox \else \vcenter \fi\fi
  \bgroup \let\par\relax
  \let\@sharp##\let\protect\relax
  \@arrayskip\@preamble}
\def\eqnarray{\stepcounter{equation}%
              \let\@currentlabel=\theequation
              \global\@eqnswtrue
              \global\@eqcnt\z@
              \tabskip\@centering                      
              \let\\=\@eqncr
              $$%
            \halign to \displaywidth  \bgroup
             \eqnumphantom \@eqnsel
      \hskip\@centering                               
    $\displaystyle  \tabskip\z@ {##}$%
    &\global\@eqcnt\@ne \hskip 2\arraycolsep
         $ \displaystyle  \arraymode{##}$\hfil
    &\global\@eqcnt\tw@ \hskip 2\arraycolsep
         $\displaystyle\tabskip\z@{##}$\hfil
         \tabskip\@centering
    &{##}\tabskip\z@\cr}
\makeatother

\def\bea{\begin{eqnarray}}
\def\eea{\end{eqnarray}}

\def\beq{\begin{equation}}
\def\eeq{\end{equation}}
\def\be{\beq\new\begin{array}{c}}
\def\ee{\end{array}\eeq}
\def\bse{\begin{subequations}}                
\def\ese{\end{subequations}}                 %

\begin{document}
\vspace{0.2cm}
\begin{center}
{\LARGE \bf A novel renormalizable representation } \\
\vspace{0.5cm}{\LARGE \bf of the Yang-Mills theory} \\
\vspace{0.7cm} {\large\bf Andrei Yu. Dubin\footnote{e-mail: dubin@itep.ru}}\\
\vspace{0.5cm}{\bf ITEP, B.Cheremushkinskaya 25, Moscow 117259, Russia}\\
 \end{center}
\vspace{0.3cm}

\begin{abstract}

For a generic gauge-invariant correlator
$<{\cal Q}[A_{\mu}]>_{A}$, we reformulate the standard $D=4$
Yang-Mills theory as a renormalizable system of two 
interacting fields $a_{\mu}$ and $B_{\mu}$ which faithfully represent high-
and low-energy degrees of freedom of the single gauge field $A_{\mu}$ in
the original formulation. It opens a possibility to synthesize an
{\it infrared-nonsingular} weak-coupling series, employed to integrate over
$a_{\mu}$ for a given background $B_{\mu}$, with qualitatively different
methods. These methods are to be applied to evaluate the resulting (after the
$a_{\mu}-$integration) representation of $<{\cal Q}[A_{\mu}]>_{A}$ in terms
of gauge-invariant generically non-local low-energy observables, like Wilson
loops. The latter observables are averaged over $B_{\mu}$ with respect to a
{\it gauge-invariant} Wilsonean
effective action $S_{eff}[B]$. To avoid a destructive dissipation between
the high- and low-energy excitations, we implement
a specific fine-tuning of the interaction between the pair of the fields:
prior to the integration over $B_{\mu}$, the expectation value
$<a_{\mu}>_{a}$ vanishes, in the tree order of the loop-wise
expansion, for an arbitrary configuration of $B_{\mu}$.

\end{abstract}

\begin{center}
\vspace{0.5cm}{Keywords: Yang-Mills, Wilsonean action, gauge invariance}\\
\end{center}

\newpage

\section{Introduction}

The $D=4$ dimensional $SU(N)$ Yang-Mills theory ($YM_{D}$), defined by the
action
\be
S_{YM}[A_{\mu}]=
\int d^{D}x~tr\left(F^{2}_{\mu\nu}(A)\right)/4g^{2}~,
\label{1.1}
\ee
belongs to a class of the systems where the physics at short distances,
characterized by asymptotic freedom, is qualitatively different compared to
the low-energy physics governed by confinement. Therefore, it
is reasonable to search for a formalism which allows to effectively combine
different computational techniques applied respectively to the high- and
low-energy dynamics interpolated at a scale $\Lambda_{int}$
sufficiently larger than\footnote{In the theory (\ref{1.1}),
the perturbative solution $g^{2}_{r}(\Lambda/\Lambda_{YM})$ of the
renormgroup equation blows up at the scale $\Lambda=\Lambda_{YM}$.}
$\Lambda_{YM}$.

The good old weak-coupling series are known to be well-defined only in the
ultraviolet ($UV$) domain of relative distances
sufficiently smaller than $\Lambda^{-1}_{YM}$.
In the $UV$ domain, the series can be extended including an input of the
infrared ($IR$) dynamics of the system. For this purpose,
the only theoretical method so far is the operator product expansion ($OPE$).
Various implementations of $OPE$ synthesize the weak-coupling
series with matrix elements of local operators which parameterize the
$IR$ input in question. Unfortunately, as well as the series itself, the
language of local operators is not robust enough to successfully
apply this method to processes dominated by the large-distance phenomena
like confinement implying a string-like pattern of the excitations.
It calls to push the idea of the synthesis even further so that the
weak-coupling expansion, in effect being restricted to the
description of the short-distance physics,
is properly combined with a description of the
$IR$ phenomena by gauge-invariant {\it non-local} correlators like Wilson
loops.

The aim of the present paper is to propose, in the Euclidean space, such a
formalism where the latter correlators are averaged directly with respect to
a gauge-invariant {\it Wilsonean} effective action which 
describes a strongly coupled gauge system representing the low-energy
dynamics of the theory (\ref{1.1}). For
this purpose, we reformulate the theory (\ref{1.1}) as a 
system of two interacting fields $a_{\mu}$ and $B_{\mu}$ which, being
described by a certain auxiliary action
$\tilde{S}_{\Lambda_{\varepsilon}}[a_{\mu},B_{\mu}]$ {\it renormalizable}
from the power counting viewpoint, represent the high- and low-energy modes of
$A_{\mu}$ respectively.

To accomplish a generic reformulation of any given gauge-invariant correlator
$<{\cal Q}[A_{\mu}]>_{A}$ in the theory (\ref{1.1}) in terms
of a pair of fields, we introduce 
a judicious Faddeev-Popov unity as a functional which depends on a
dynamical field $B_{\mu}$. The corresponding gauge condition is imposed on
the combination $A_{\mu}-B_{\mu}$ to be identified with the field $a_{\mu}$.
As a result, $<{\cal Q}[A_{\mu}]>_{A}$ is rewritten in the form
\be
\Big<{\cal Q}[A_{\mu}]\Big>_{A}=
\Big<~~\Big<{\cal Q}[a_{\mu}+B_{\mu}]\Big>_{a}^{B^{ext}}\Big>_{B}~,
\label{15.14}
\ee                   
where the averaging $<..>_{a}^{B}$ over the high-energy field $a_{\mu}$
is performed, for a given low-energy field
$B_{\mu}\equiv B^{ext}_{\mu}$ considered as
external, with the "microscopic" action
$\tilde{S}_{\Lambda_{\varepsilon}}[a_{\mu},B_{\mu}]$.
It allows to compute both the various
averages $<..>^{B}_{a}$ and the associated partition function $Z_{1}[B]$ (of
the auxiliary high-energy theory for a fixed $B_{\mu}$) using the $1/N$
weak-coupling series running in the renormalized
coupling constant $g^{2}_{r}\equiv g^{2}_{r}(\Lambda/\Lambda_{YM})$ associated with a scale
$\Lambda$. In particular, to avoid
a destructive dissipation between the high- and low-energy
excitations, the interaction between the pair of the fields has to be
judiciously constrained. We impose that, at least in the tree-order of the
renormalized loop-wise expansion in the external background
$B_{\mu}\equiv B^{ext}_{\mu}$, the constraint
\be
\Big<a_{\mu}({\bf x})\Big>_{a}^{B^{ext}}=0~~~~~,~~~~~\forall{B_{\mu}}~,
\label{RS.01}
\ee
holds true for a {\it generic} configuration of $B_{\mu}$.
In the high-energy sector, it is shown to maintain that the spurious
'symmetry breaking' (displayed by $<a_{\mu}({\bf x})>_{a}^{B}\neq 0$), being
suppressed by powers of the coupling constant
$g^{2}_{r}$, leaves the background perturbation theory well-defined.

The subsequent integration $<..>_{B}$
over the low-energy field $B_{\mu}$ is performed with respect to the
corresponding effective action $S_{eff}[B_{\mu}]$
conventionally defined by the relation
\be
exp\left(-S_{eff}[B_{\mu}]\right)=Z_{1}[B]~,
\label{15.16}
\ee                   
where the partition function $Z_{1}[B]$ is introduced above. To maintain a
gauge-invariant description of the low-energy phenomena, the action
$\tilde{S}_{\Lambda_{\varepsilon}}[a_{\mu},B_{\mu}]$ is imposed to be 
invariant under the background gauge transformations
\cite{BF,Abb}:
\be
B_{\mu}~\longrightarrow{~B_{\mu}^{(\psi)}=
U(\psi)(B_{\mu}+i\partial_{\mu})U^{-1}(\psi)}~~~~~,~~~~~a_{\mu}~
\longrightarrow{~a_{\mu}^{(\psi)}=U(\psi)a_{\mu}U^{-1}(\psi)}~,
\label{15.6}
\ee
which entails that the associated effective action
$S_{eff}[B]$ respects the gauge symmetry. In turn, it implies that, after the
integration over $a_{\mu}$, the average (\ref{15.14}) is indeed expressed
in terms of (non-local) gauge-invariant correlators depending on $B_{\mu}$.
Although it is definitely beyond the scope of the paper to discuss a scheme
for evaluation of the low-energy correlators $<..>_{B}$, we note that
these correlators may be approached employing the stringy form of the $1/N$
strong-coupling expansion \cite{Dub3} (see also \cite{Dub&Mak}) yielding a
continuum counterpart of the corresponding lattice expansion.
In this way, the reformulation (\ref{15.14}) of (\ref{1.1}) is suggested to
provide a bridge to interpolate between the $1/N$ weak- and $1/N$ 
strong-coupling series.

Next, the resolution of the constraint (\ref{RS.01}), truncated to a given
order of the loop-wise expansion, is to be understood
in the context of the following prescription that takes advantage of
the freedom in the choice of
$\tilde{S}_{\Lambda_{\varepsilon}}[a_{\mu},B_{\mu}]$.
To begin with, it is convenient to impose that the transformation
(\ref{15.14}) results in the axial gauge condition for $a_{\mu}$:
\be
f^{c}(a,B)=n_{\mu}a^{c}_{\mu}=0~~~~~,~~~~~~n^{2}_{\mu}=1~,
\label{15.14i}
\ee
where $n_{\mu}$ is a constant $D-$vector, and we presume that
$n_{\mu}a_{\mu}=a_{0}=0$ which leaves $D-1$ dynamical components $a_{i}$.
In addition to Eq. (\ref{15.14i}), we impose
that the difference
$\tilde{S}_{\Lambda_{\varepsilon}}[a_{i},B_{\mu}]-
{S}_{YM}[a_{\mu}+B_{\mu}]|_{a_{0}=0}$ defines a Lagrangian which is a
quadratic polynomial in $a_{i}$ with generically $B_{\mu}-$dependent
coefficients. The quadratic in $a_{i}$ term serves merely to attribute, at
the tree-level, a mass ($\sim\Lambda_{int}$)
to the $a_{i}-$field which, in turn, facilitates the implementation of
the transformation (\ref{15.14}) as a multiscale decomposition.
As a by-product, for sufficiently large value of the interpolation scale
$\Lambda_{int}$, 
the background perturbation theory
is free of spurious $IR$ divergences. It is also noteworthy that, akin to
the case of the original theory (\ref{1.1}), there are only {\it two}
propagating polarizations of the field $a_{i}$ (as it is formalized by eq.
(\ref{TR.05})). Concerning the linear in
$a_{i}$ term, it is completely determined by the constraint (\ref{RS.01}) or
its $L-$loop truncation.
E.g., in the leading tree-order of the loop-wise expansion,
Eq. (\ref{RS.01}) reduces to the requirement that, for {\it any} $B_{\mu}$,
the tree-level approximation\footnote{This approximation is conventionally
obtained from the renormalized representation
$\tilde{S}^{r}_{\Lambda_{\varepsilon}}[a_{i},B_{\mu}]$
of $\tilde{S}_{\Lambda_{\varepsilon}}[a_{i},B_{\mu}]$ after the exclusion of
the relevant counterterms.} $\tilde{S}^{tr}[a_{i},B_{\mu}]$
to $\tilde{S}_{\Lambda_{\varepsilon}}[a_{i},B_{\mu}]$
does {\it not} contain a term linear in $a_{i}$ that otherwise would make the
renormalized background perturbation theory ill-defined. Then, Eq.
(\ref{15.14}) yields a unique way to fix the remaining $a_{i}-$independent
part of the above difference. At least when Eq. (\ref{RS.01}) is truncated
to the leading $L=0$ order, the resulting
$\tilde{S}_{\Lambda_{\varepsilon}}[a_{i},B_{\mu}]$
complies with the renormalizability from the power counting viewpoint,
provided a pair of auxiliary ghost-fields is introduced.

Finally, once the residual symmetry (\ref{15.6}) is fixed, thus implemented
transformation (\ref{15.14}) should yield such realization of the multi-scale
decomposition of the theory (\ref{1.1}) that the effective action
(\ref{15.16}) is indeed of the Wilsonean type.
Qualitatively, the action
$\tilde{S}_{\Lambda_{\varepsilon}}[a_{\mu},B_{\mu}]$ should enforce that the
Fourier images of $a_{\mu}$ and $B_{\mu}$ are dynamically localized in (but
generically {\it not} limited to) the corresponding momentum slices
$[\Lambda_{int},\Lambda_{\varepsilon}]$ and $[0,\Lambda_{int}]$. For this
purpose, the interpolation scale $\Lambda_{int}$ is to be identified with the
$IR$ limit ${\cal M}_{a}$ (see eq. (\ref{CT.01k})) of the renormalized mass of
the $a_{i}-$field in the auxiliary high-energy theory defining the correlators
$<..>_{a}^{B}$, while the limit
$ln(\Lambda_{\varepsilon}/\Lambda_{int})\sim{1/\varepsilon}
\rightarrow{\infty}$ is maintained via the dimensional
regularization with $\varepsilon=4-D\rightarrow{+0}$.
In due course, we demonstrate that the proposed below action
$\tilde{S}_{\Lambda_{\varepsilon}}[\cdot]$,
being conventionally renormalizable, satisfies certain precise conditions
(\ref{CT.01v}) which do imply the Wilsonean type of $S_{eff}[B_{\mu}]$.
Also, to make the proposed perturbative computation of the
effective action (\ref{15.14}) tractable, it is
important to choose such renormalization scale $\Lambda=\breve{\cal M}$
(implicitly entering, via $g^{2}_{r}(\Lambda/\Lambda_{YM})$, the definition
of $<a_{\mu}({\bf x})>_{a}^{B}$ in
eq. (\ref{RS.01})) that is judiciously adjusted to
the interpolation scale
$\Lambda_{int}={\cal M}_{a}$ according to eq. (\ref{CT.01kk}).
Altogether, thus implemented eq. (\ref{15.14}) generalizes the
transformation\footnote{Utilizing this transformation only once after a
gauge fixing in the theory (\ref{1.1}), one obtains the action
$\tilde{S}_{\Lambda_{\varepsilon}}[a_{\mu},B_{\mu}]$ which
differs from ${S}_{YM}[a_{\mu}+B_{\mu}]$ only by kinetic terms quadratic in
$a_{\mu}$ and $B_{\mu}$. In turn, it allows to fulfil neither the
symmetry (\ref{15.6}) nor (even the tree-order approximation) the
condition (\ref{RS.01}).}
\cite{RG} that, after infinitely many
applications of its
$(\Lambda_{\varepsilon}-\Lambda_{int})/\Lambda_{\varepsilon}\rightarrow{+0}$
version for a fixed $\Lambda_{\varepsilon}$, facilitates
modern approaches to perform the renormgroup reduction
of the high-momentum Fourier modes of a given quantum field.


In Section \ref{DAG}, we introduce the relevant variety of the transformations
(\ref{15.14}) parameterized by a single function ${\cal T}_{i}(B)$. It is
done in the simplest setting when the axial gauge $A_{\mu}n_{\mu}=A_{0}=0$
is fixed prior to the transformation. It results in the
theory of the two fields $a_{\mu}$ and $B_{\mu}$ where, in addition to the
condition (\ref{15.14i}), the residual invariance (\ref{15.6}) is also fixed
by the second gauge condition $B_{\mu}n_{\mu}=0$.
In Section \ref{restoration}, synthesizing the latter variant of the
transformation with
a gauge fixing unity, we generalize the construction so that, keeping
the symmetry (\ref{15.6}) manifest, the decomposition is
performed for a general class of gauge conditions for $a_{\mu}$.
Also, we comment on the case when the auxiliary action
$\tilde{S}_{\Lambda_{\varepsilon}}[a_{\mu},B_{\mu}]$ maintains a
generic $a_{\mu}-$independent difference
$\tilde{S}_{\Lambda_{\varepsilon}}[a_{\mu},B_{\mu}]-S_{YM}[a_{\mu}+B_{\mu}]$
so that the transformation (\ref{15.14}) reduces to an identity
(attributed to 't Hooft) used in an approach \cite{BPT}.

In Section \ref{resolution}, the condition (\ref{RS.01}) is
reformulated as a simple algebraic equation that can be used to
unambiguously determine the function ${\cal T}_{i}(B)$ order by order in
the framework of the renormalized loop-wise expansion applied prior to the
averaging over $B_{\mu}$.
The explicit form of ${\cal T}_{i}(B)$ is obtained in the tree-order of this
expansion, while the renormalizability of the resulting theory is sketched in
Section \ref{renorm}, where the structure of the counterterms is also
discussed. In Section \ref{counterterms}, 
thus implemented transformation (\ref{15.14}) is shown to guarantee
the Wilsonean type of the effective action (\ref{15.16}).

\section{The general trick in the gauge $a_{0}=B_{0}=0$}
\label{DAG}

The short-cut route to a transformation (\ref{15.14}) consistent with
the symmetry (\ref{15.6}) is to implement the transformation after the
gauge fixing $A_{0}=0$ so that the resulting action
$\tilde{S}_{\Lambda_{\varepsilon}}[a_{i},B_{i}]$ forms the $B_{0}=0$
reduction of a functional invariant under
(\ref{15.6}). The proposal is to first find such an action
\be
{S}_{m}[a_{i},B_{\mu}]=
\left(\tilde{S}_{\Lambda_{\varepsilon}}[a_{\mu},B_{\mu}]-
{S}_{YM}[a_{\mu}+B_{\mu}]\right)\Big|_{a_{0}=0}
\label{AC.01}
\ee                   
which resolves the condition that
\be
1=\int {\cal D}B_{i}~
exp\left(-{S}_{m}[A_{i}-B_{i},B_{\mu}]\right)\Big|_{B_{0}=A_{0}}~,
\label{RE.05b}
\ee                   
is fulfilled for an {\it arbitrary} $A_{\mu}({\bf x})=(A_{0}({\bf x}),
A_{i}({\bf x}))$. Then, one is to insert $B_{0}=A_{0}=0$ option of the the
unity (\ref{RE.05b}) under the axial gauge $A_{0}=0$ implementation of
the generating functional
\be
\Big<{\cal Q}[A_{\mu}]\Big>_{A}=\frac{1}{Z_{YM}}\int
\frac{{\cal D}A_{\mu}}{{\cal D} \omega}~
exp\left(-S_{YM}[A_{\mu}]\right)~{\cal Q}[A_{\mu}]~,
\label{GI.04}
\ee
where $S_{YM}[A_{\mu}]$ is given in eq. (\ref{1.1}), ${\cal Q}[A_{\nu}]$
parameterizes a generic gauge-invariant external source, 
$Z_{YM}$ denotes the partition function of the Euclidean gauge theory
(\ref{GI.04}), and the measure
${\cal D}A_{\mu}/{\cal D} \omega$ includes the normalization factor to cancel
the volume $\int {\cal D} \omega$ of the group of the standard gauge
transformations. The reformulation (\ref{15.14}) is completed through the
subsequent change of the pair of the variables
${\cal D}B_{i}{\cal D}A_{i}\rightarrow{{\cal D}B_{i}{\cal D}a_{i}}$,
$a_{i}+B_{i}=A_{i}$.
Altogether, it results in such decomposition\footnote{It
reduces to the one of \cite{RG} provided $2g^{2}_{r}{S}_{m}[a_{i},B_{i}]=
\int d^{D}x~Tr(a_{i}\hat{\cal K}^{ij}_{1}a_{j}+
B_{i}\hat{\cal K}^{ij}_{2}B_{j}-(a_{i}+B_{i})\hat{\cal K}^{ij}(a_{j}+
B_{j}))$, where $\hat{\cal K}^{ij}$ is defined after eq. (\ref{TR.05})
and $\hat{\cal K}^{ij}_{k}$ are generic operators satisfying the condition
$\hat{\cal K}^{-1}=\hat{\cal K}_{1}^{-1}+\hat{\cal K}_{2}^{-1}$.}
(\ref{15.14}) where the integration over the high-energy modes
$a_{\mu}$ is performed in compliance with the $B_{0}=0$ variant of the
$a_{\mu}n_{\mu}=a_{0}$ prescription:
\be
\Big<{\cal Q}[a_{\mu}+B_{\mu}]\Big>_{a}^{B^{ext}}=
\frac{1}{Z_{1}[B]}
\int{\cal D}a_{\mu}~\delta\left(n_{\mu}a_{\mu}\right)~
{\cal Q}[a_{\mu}+B_{\mu}]
~exp\left(-\tilde{S}_{\Lambda_{\varepsilon}}[a_{\mu},B_{\mu}]\right)~,
\label{GI.06}
\ee
and the relevant microscopic action
$\tilde{S}_{\Lambda_{\varepsilon}}[a_{i},B_{\mu}]$
is given by eq. (\ref{AC.01}), while the intermediate partition function
(\ref{15.16}) is such that $<1>_{a}^{B^{ext}}=1$.

Finally, it is straightforward to maintain that, in addition to
(\ref{15.14i}), the above decomposition indeed implies one more axial gauge
fixing associated with the transformations (\ref{15.6}). For this purpose,
it is sufficient to impose that
${S}_{m}[a_{i},B_{\mu}]$ is invariant under (\ref{15.6}),
\be
{S}_{m}[a_{i},B_{\mu}]=
{S}_{m}[a_{i}^{(\psi)},B^{(\psi)}_{\mu}]~\Longrightarrow~
\tilde{S}_{\Lambda_{\varepsilon}}[a_{i},B_{\mu}]=
\tilde{S}_{\Lambda_{\varepsilon}}[a_{i}^{(\psi)},B^{(\psi)}_{\mu}]~,
\label{15.12v}
\ee
which entails, due to eq. (\ref{AC.01}), the same invariance of the full
action defined $\tilde{S}_{\Lambda_{\varepsilon}}[\cdot]$.
Then, the remaining averaging over the low-energy modes can be reformulated 
as the $B_{0}=0$ gauge implementation of the prescription which,
similarly to eq. (\ref{GI.04}), manifestly respects gauge symmetry:
\be
\Big<{\cal G}[B_{\mu}]\Big>_{B}=\frac{1}{Z_{YM}}\int
\frac{{\cal D}B_{\mu}}{{\cal D} \psi}~{\cal G}[B_{\mu}]~
e^{-S_{eff}[B_{\mu}]}~~~~~,~~~~~S_{eff}[B_{\mu}]=S_{eff}[B^{(\omega)}_{\mu}]
\label{GI.03}
\ee
where $Z_{YM}$ is the same as in (\ref{GI.04}), $<1>_{B}=1$, and
the effective action (\ref{15.16}) is gauge-invariant.

\subsection{The ansatz for ${S}_{m}[a_{i},B_{\mu}]$}

To resolve the constraint (\ref{RS.01}) in the framework
of the renormalized background perturbation theory,
we propose to resolve the condition (\ref{RE.05b}) by the
${\cal T}_{i}(B)-$dependent ansatz
\be
S_{m}[a_{i},B_{\mu}]={\cal X}[w_{j}(a_{i},B_{\mu})]-
ln\left(det\left[\hat{\cal E}_{ij}(B)\right]\right)+
ln({\cal Z}_{{\cal X}})~,
\label{MR.02}
\ee
where
$B_{\mu}$ denotes the {\it full} $D-$vector, and
\be
{\cal X}[w_{i}]=
\int d^{D}x~\frac{{\cal M}^{2}}{2g^{2}_{r}}~Tr\left(w^{2}_{i}
\right)~~~,~~~w_{i}(a_{j},B_{\mu})=-a_{i}-
\frac{g^{2}_{r}}{{\cal M}^{2}}{\cal T}_{i}(B_{\mu})~,
\label{RE.08}
\ee
and $g^{2}_{r}\equiv g^{2}_{r}(\Lambda/\Lambda_{YM})$ denotes
the renormalized coupling constant $g^{2}/Z_{g^{2}}$ in the
original formulation (\ref{1.1}) of the theory which is associated with a
finite, when $4-D\rightarrow{+0}$, normalization point $\Lambda$. Also,
it is convenient to choose
such value $\breve{\cal M}$ of $\Lambda$ that the interpolation
scale $\Lambda_{int}={\cal M}_{a}$ coincides with the parameter ${\cal M}$
(as it will be formalized by eq. (\ref{CT.01kk})).



As for $det[\cdot]$, being evaluated with respect to both pairs of the
indices of $\hat{\cal E}^{cd}_{ij}[B]$, it yields the Jacobian
associated with the change of the variables $B_{i}
\rightarrow{w_{i}(A_{j}-B_{j},B_{\mu})}$ performed for a fixed
$B_{0}$. The corresponding tensor-like
operator $\hat{\cal E}^{cd}_{\mu\nu}(B)$ is therefore defined via the relation
\be
<{\bf y}|\left(\hat{\cal E}_{ij}^{cd}(B)-
\delta_{ij}\delta^{cd}\right)|{\bf x}>=-
\frac{g^{2}_{r}}{{\cal M}^{2}}
\frac{\delta {\cal T}_{i}^{c}(B({\bf y}))}
{\delta B_{j}^{d}({\bf x})}~,
\label{RE.09}
\ee
and ${\cal T}_{\mu}(\cdot)$ should depend only
on the multiplicatively renormalized quantities which are finite
in the limit $\varepsilon=4-D\rightarrow{+0}$. Finally, the constant
${\cal Z}_{{\cal X}}$ is defined
by the relation ${\cal Z}_{{\cal X}}=\int {\cal D}w_{i}~
e^{-{\cal X}^{r}[w_{i}]}$ which ensures that, after the above change of the
variables, 
the ansatz (\ref{MR.02}) indeed resolves the condition\footnote{In eq.
(\ref{RE.05b}), the functional measure
is presumed to be defined so that ${\cal Z}_{{\cal X}}$ is finite when
$4-D\rightarrow{+0}$.} (\ref{RE.05b}).
In turn, the required invariance
(\ref{15.12v}) is evidently maintained provided 
\be
{\cal T}_{j}(B^{(\psi)})=U(\psi){\cal T}_{j}(B)U^{-1}(\psi)~,
\label{CV.01}
\ee                   
i.e., both the function ${\cal T}_{j}(B({\bf z}))$ and,
in consequence, the operator $\hat{\cal E}_{ij}(B)$ are transformed 
{\it covariantly} under the ordinary gauge symmetry.

Next, the high-energy averages (\ref{GI.06}) are, by construction, invariant
under the
replacement of $\tilde{S}^{r}_{\Lambda_{\varepsilon}}[a_{i},B_{\mu}]$ by the
simpler action (\ref{RS.08}).
\be
\bar{S}^{r}[a_{i},B_{\mu}]=\breve{S}^{r}[a_{i},B_{\mu}]+
\int d^{D}x~Tr(a_{j}{\cal T}_{j}(B))
\label{RS.08}
\ee
resulting when the last two $a_{i}-${\it independent} terms in eq.
(\ref{MR.02}) are omitted so that
\be
\breve{S}^{r}[a_{i},B_{\mu}]=S^{r}_{YM}[a_{\mu}+B_{\mu}]\Big|_{a_{0}=0}+
\int d^{D}x~{{\cal M}^{2}}~Tr\left(a^{2}_{i}\right)/{2g^{2}_{r}}~,
\label{RS.09}
\ee
where $S^{r}_{YM}[A_{\mu}]$ is obtained from
$S_{YM}[A_{\mu}]$ rewriting $g^{2}=Z_{g^{2}}g^{2}_{r}$.
We utilize that the auxiliary theory (\ref{RS.08}), considered for a
fixed $B_{\mu}$ the axial gauge fixing for $a_{\mu}$, is conventionally
renormalizable. It is also noteworthy that, according to the conditions
(\ref{CT.01v}), in the limit $\varepsilon\rightarrow{+0}$ neither the
parameter ${\cal M}$ nor the involved gauge fields require a multiplicative
renormalization in the framework of the background perturbation theory:
$a_{i}^{r}=a_{i}$, $B_{\mu}^{r}=B_{\mu}$, ${\cal M}^{r}={\cal M}$. 
The conditions (\ref{CT.01v}) also imply that the renormalization of the
entire gauge system {\it reduces} to the one of the 
perturbative expansion in the theory (\ref{RS.08}) of
the single dynamical field $a_{i}$. In particular, for a given normalization
point $\Lambda$,
the renormalization of $g^{2}$ in the latter theory is maintained via the 
same factor $Z_{g^{2}}$ 
as in the original formulation (\ref{1.1}) considered in the gauge $A_{0}=0$.


Finally, for our later purposes, we introduce vector-like ghost
fields $\bar{\vartheta}_{i}$ and ${\vartheta}_{i}$ according to
the representation
\be
det\left[\hat{\cal K}_{ij}(B)\right]=\int {\cal D}\bar{\vartheta}_{i}({\bf z})
{\cal D}{\vartheta}_{i}({\bf z})~exp\left[\int d^{D}x~Tr\left(
\bar{\vartheta}_{i}~\hat{\cal K}_{ij}(B)~{\vartheta}_{j}\right)\right]~.
\label{FM.01m}
\ee
which, in effect, replaces $ln(det[\cdot])$ in eq. (\ref{MR.02}) by the
functional in the exponent in the r.h. side of eq. (\ref{FM.01m}) so that
$det[\hat{1}_{ij}]=1$. Let us stress that the above massive fermionic fields
$\bar{\vartheta}_{i}$, ${\vartheta}_{i}$ should {\it not} be interpreted
as some extra high-energy modes additional to $a_{i}$. Indeed,
as the high-energy averages (\ref{GI.06}) are defined by the
action (\ref{RS.08}),  the second term in the r.h. side of eq. (\ref{MR.02})
enters the decomposition (\ref{15.14}) only as the associated part
of the low-energy effective
action $S_{eff}[B_{\mu}]$ entering the average (\ref{GI.03}).

\section{Restoration of the explicit background gauge invariance}
\label{restoration}

Actually, the condition (\ref{RE.05b}) can be generalized to
implement a generic gauge fixing for the field $a_{\mu}$ keeping the
background gauge invariance (\ref{15.6}) manifest. Given the generalized
construction (see eq. (\ref{15.13}) below), eq. (\ref{RE.05b}) is reproduced
imposing in ${\bf E}^{4}$ the double axial gauge
$n_{\mu}a^{b}_{\mu}=n_{\mu}B^{b}_{\mu}=0$ in the two successive steps so that
the invariance (\ref{15.6}) is fixed only in the very end.
For simplicity, we restrict our attention to
the subvariety of the {\it linear} background gauges
$f^{c}(a,B)={\cal R}^{ce}_{\mu}(B)a^{e}_{\mu}=0$
where, in order to maintain the required symmetry of 
$\tilde{S}_{\Lambda_{\varepsilon}}[a,B]$, the operator ${\cal R}_{\mu}(B)$ is
constrained to transform homogeneously under
the transformations (\ref{15.6}): ${\cal R}_{\mu}(B^{(\psi)})=
U(\psi){\cal R}_{\mu}(B)U^{-1}(\psi)$.

The form of the multi-scale decomposition, respecting the latter symmetry,
can be introduced judiciously synthesizing a transformation
like (\ref{15.14}) with the Faddeev-Popov unity adapted
to fix, in accordance with (\ref{15.6}), a gauge for the high-energy field
$a_{\mu}$ represented by the combination $A_{\mu}-B_{\mu}$.
The proposal is to utilize the following $B_{\mu}-$dependent functional
\be
1=
~\int \frac{{\cal D}B_{\mu}}{{\cal D}\psi}~
exp\left(-\tilde{S}_{m}[A^{(\omega_{0}[B])}_{\mu}-B_{\mu},B_{\mu}]
\right)
\label{15.13}
\ee                   
as the composed unity, where
$A^{(\omega)}_{\mu}=U(\omega)(A_{\mu}+i\partial_{\mu})U^{-1}(\omega)$ and
the auxiliary action
$\tilde{S}_{m}[a_{\mu},B_{\mu}]$, being invariant under (\ref{15.6}), is
such that the condition (\ref{15.13}) holds true for $\forall{A_{\mu}}$.
Implying the necessity of the factor $1/{\cal D}\psi$, the functional
$\omega_{0}[B]\equiv\omega_{0}[A_{\mu},B_{\mu}]$ is
determined by the relation
\be
e^{-\tilde{S}_{m}[A^{(\omega_{0}[B])}_{\mu}-B_{\mu},B_{\mu}]}
=\int {\cal D}\omega~
det\left[f'(A^{({\omega})}_{\mu}-B_{\mu},B_{\mu})\right]
~\delta\left(f(A^{({\omega})}_{\mu}-B_{\mu},B_{\mu})\right)
e^{-\tilde{S}_{m}[A^{({\omega})}_{\mu}-B_{\mu},B_{\mu}]}~.
\label{15.13x}
\ee                   
where $f'(\cdot)\equiv\delta f(\cdot)/\delta {\omega}$, and the
shift ${\omega}\rightarrow{\omega}\circ\psi$ reveals that the r.h.
side is invariant under the
$a_{\mu}\rightarrow{A^{({\omega})}_{\mu}-B_{\mu}}$ option of the
transformations
(\ref{15.6}) with $A^{({\omega}\circ\psi)}_{\mu}-B^{(\psi)}_{\mu}
=U(\psi)(A^{({\omega})}_{\mu}-B_{\mu})U^{-1}(\psi)$.
In turn, it implies that $\omega_{0}[A_{\mu},B^{(\psi)}_{\mu}]=
\omega_{0}[A_{\mu},B_{\mu}]\circ\psi$ which explains the necessity
to cancel in eq. (\ref{15.13}) the volume
$\int {\cal D} \psi$ of the group of the transformations (\ref{15.6}).

Then, akin to the previous Section, one is to insert the unity (\ref{15.13})
under the functional integral (\ref{GI.04}) and, after simple manipulations,
we arrive at the relation (\ref{15.14}). Its particular form is specified by
eqs. (\ref{GI.06}) and (\ref{GI.03}), provided the identification
\be
{S}_{m}[a_{\mu},B_{\mu}]=\tilde{S}_{m}[a_{\mu},B_{\mu}]-
ln\left(det\left[{\cal R}_{\mu}(B)D_{\mu}(a+B)\right]\right)~,
\label{15.16a}
\ee                   
is made in the definition (\ref{AC.01}) of
$\tilde{S}_{\Lambda_{\varepsilon}}[a_{\mu},B_{\mu}]$, while $n_{\mu}$ is
replaced by ${\cal R}_{\mu}(B)$ in eq. (\ref{GI.06}).
In consequence, employing that
${\cal R}_{\mu}(B^{(\psi)})=
U(\psi){\cal R}_{\mu}(B)U^{-1}(\psi)$,
the condition (\ref{15.12v}) is indeed sufficient
to maintain the background gauge invariance (\ref{15.6}) of
thus introduced action $\tilde{S}_{\Lambda_{\varepsilon}}[a_{\mu},B_{\mu}]$.
In turn, it allows to rewrite the condition (\ref{15.13}) in the form:
\be
1=~\int {\cal D}B_{\mu}~\det\left[{\cal R}_{\mu}(B)D_{\mu}(A)\right]
\delta\left({\cal R}_{\mu}(B)(A_{\mu}-B_{\mu})\right)
exp\left(-\tilde{S}_{m}[A_{\mu}-B_{\mu},B_{\mu}]\right)~,
\label{CS.01}
\ee                   
In the axial gauge (\ref{15.14i}), integrating over the longitudinal
component $n_{\mu}B_{\mu}$ of $B_{\mu}$, one reduces the condition
(\ref{CS.01}) to the constraint (\ref{RE.05b}).

Finally, we remark that, when $\tilde{S}_{m}[a_{\mu},B_{\mu}]=
\tilde{S}_{m}[B_{\mu}]$ is $a_{\mu}-$independent, the insertion
of the unity (\ref{15.13}) does not impose any gauge
fixing for $a_{\mu}=A_{\mu}-B_{\mu}$ which can be
performed subsequently.
Thus reduced unity (\ref{15.13}) yields the transformation of the
generation functional (\ref{GI.04}) which reproduces the so-called 't Hooft
identity that, in \cite{BPT}, is claimed  (without a specification of
$\tilde{S}_{m}[B_{\mu}]$) to help in separation of confining
$B_{\mu}-$configurations. Irrespectively of a choice of
$\tilde{S}_{m}[B_{\mu}]$, such a transformation is ineffective
to implement a multi-scale decomposition: it does {\it not}
attribute a mass term to the field $a_{\mu}$. Consequently, prior to the
integration over $B_{\mu}$, the contribution of
the low-energy modes of $a_{\mu}$ is unsuppressed, and 
the effective action (\ref{15.16}) is not of the Wilsonean type. Also, for
$\forall{\tilde{S}_{m}[B_{\mu}]}$, the
condition (\ref{RS.01}) is violated already at the tree-level of the
loop-wise expansion.

\section{Resolving the constraint (\ref{RS.01})}
\label{resolution}

To demonstrate that the condition (\ref{RS.01}) unambiguously determines the
function ${\cal T}_{\mu}(B)$ entering the ansatz (\ref{RE.08}), we
begin with the following observation. To begin with, presuming
$n_{\mu}a_{\mu}=a_{0}$, eq. (\ref{RS.01}) can
be rewritten as the constraint ${\delta
{\cal W}^{r}[J_{i}|B_{\mu}]}/{\delta J_{j}({\bf z})}|_{J_{i}=0}=0$. Here,
${\cal W}^{r}[\cdot]$ denotes the relevant renormalized generating functional
expressed in terms of the coupling constant
$g^{2}_{r}=g^{2}/Z_{g^{2}}$ and the fields $a_{i}=a^{r}_{i}$,
$B_{\mu}=B^{r}_{\mu}$ renormalized according to the discussion after
eq. (\ref{RS.09}). This functional is defined by the relation
$e^{{\cal W}^{r}[J_{i}|B_{\mu}]}=
<e^{\int d^{D}x~Tr(a_{i}J_{i})}>_{a}^{B^{ext}}$,
where the high-energy quantum averaging is introduced in eq. (\ref{GI.06}).
In turn, the latter constraint can be replaced by the condition
\be
{\delta \Gamma^{r}[C_{i}|B_{\mu}]}/{\delta C_{j}({\bf z})}
\Big|_{C_{i}=0}=0
\label{RS.05}
\ee                   
formulated in terms of the {\it renormalized} Legendre effective action
$\Gamma^{r}[C_{i}|B_{\mu}]$ determined by the canonical relation:
$\Gamma^{r}[C_{i}|B_{\mu}]+{\cal W}^{r}[J_{i}|B_{\mu}]-
\int d^{D}x~Tr\left(J_{i}C_{i}\right)=0$, where
$J^{r}_{i}=J_{i}$ and $C_{i}^{r}=C_{i}$ in view of eq.
(\ref{RS.02}). Indeed, eq. (\ref{RS.05}) follows from the general property
that ${\delta \Gamma^{r}[C_{i}|B_{\mu}]}/{\delta C_{j}({\bf z})}=0$ for
$C_{i}({\bf x})=<a_{i}({\bf x})>_{a}^{B^{ext}}$ which is a consequence of
the above relation between $\Gamma^{r}[\cdot]$ and
${\cal W}^{r}[\cdot]$.

Next, we utilize that the constraint (\ref{RS.05}) is not changed if
$\Gamma^{r}[C_{i}|B_{\mu}]$ is replaced by the Legendre
effective action in the simpler theory with the modified action (\ref{RS.08}).
Furthermore, using the {\it linearity} of the reduced action (\ref{RS.08}) in
${\cal T}_{j}(B)$, one readily obtains that
$\Gamma^{r}[C_{i}|B_{\mu}]=\breve{\Gamma}^{r}[C_{i}|B_{\mu}]+
\int d^{D}x~Tr(C_{j}{\cal T}_{j}(B))$,
where $\breve{\Gamma}^{r}[C_{i}|B_{\mu}]$ is the renormalized Legendre
effective action associated with the generating functional defined
by the action (\ref{RS.09}). Altogether, implementing
the background perturbation theory associated with a given normalization
point $\Lambda$, eq. (\ref{RS.05}) can be rewritten in
the form
\be
{\cal T}_{j}(B({\bf z}))=-\frac{\delta \breve{\Gamma}^{r}[C_{i}|B_{\nu}]}
{\delta C_{j}({\bf z})}\Big|_{C_{i}=0}=
\sum_{l=0}^{\infty}{\cal T}_{j}^{(l)}(B)~,
\label{RS.06}
\ee                   
where the expansion of ${\cal T}_{j}(B)$ is generated by the renormalized
loop-wise expansion of ${\Gamma}^{r}[\cdot|\cdot]=
\sum_{l=0}^{\infty}[\breve{\Gamma}^{r}_{l}[\cdot|\cdot]+
\int d^{D}x~Tr(C_{j}{\cal T}_{j}^{(l)}(\cdot))]$ so that
${\cal T}_{j}^{(l)}(B)\sim{g^{2l-2}_{r}(\Lambda/\Lambda_{YM})}$. To
complete the specification of the renormalized form of the
action (\ref{15.16a}), the (truncated) pattern (\ref{RS.06}) is to be
substituted into eq. (\ref{RE.09}) defining the operator
$\hat{\cal E}_{ij}(B)=\sum_{l} \hat{\cal E}_{ij}^{(l)}(B)$.

In the tree-approximation, the ansatz
(\ref{MR.02}) is defined by $\breve{\Gamma}^{r}_{0}[C_{i}|B_{\nu}]=
\breve{S}^{tr}[C_{i},B_{\nu}]$:
\be
\left({\cal T}_{j}^{(0)}(B({\bf z}))\right)^{b}=-
\frac{\delta \breve{S}^{tr}[C_{i},B_{\nu}]}{\delta C^{b}_{j}({\bf z})}
\Big|_{C_{i}=0}=-
\frac{\delta S^{tr}_{YM}[B_{\nu}]}{\delta B^{b}_{j}({\bf z})}=
\frac{1}{g^{2}_{r}}
D^{bc}_{\mu}(B({\bf z}))F^{c}_{\mu j}(B({\bf z}))~,
\label{MR.01}
\ee
where the tree-level action $\breve{S}^{tr}[\cdot]$ is conventionally
obtained from eq. (\ref{RS.09}) replacing $S^{r}_{YM}[B_{\nu}]$ by
$S^{tr}_{YM}[B_{\nu}]$ which, in turn,
implies the replacement of $g^{2}=Z_{g^{2}}g^{2}_{r}$ by
$g^{2}_{r}\equiv{g^{2}_{r}(\Lambda/\Lambda_{YM})}$.
Correspondingly, the leading approximation to
operator $\hat{\cal E}_{ij}(B)$ reads
\be
\left(\hat{\cal E}^{(0)}_{ij}(B)\right)^{bd}-
\delta_{ij}\delta^{bd}=-\frac{1}{{\cal M}^{2}}\left(
{D}^{bc}_{\rho}(B){D}^{cd}_{\rho}(B)\delta_{ij}-
{D}^{bc}_{i}(B){D}^{cd}_{j}(B)-2f^{bde}{F}^{e}_{ij}(B)\right)~.
\label{MR.04}
\ee

Next, to evaluate the average (\ref{GI.06}) up
to a given order $L\geq{0}$ of the loop-wise expansion,
$\breve{\Gamma}^{r}[C_{i}|B_{\nu}]$ is sufficient to
determine up to the same order of the expansion. In eq. (\ref{RS.08}), in
the sum $\sum_{l=0}^{L}a_{j}{\cal T}_{j}^{(l)}(B)$, only
the term $a_{j}{\cal T}_{j}^{(0)}(B)$ is to be involved
in the derivation of the propagator $G^{r}_{ij}({\bf y},{\bf x}|B)$ of the
renormalized perturbation theory for a fixed $B_{\mu}$. Then,
according to eq. (\ref{RS.05}), (for $\forall{L}$) the
tree-level approximation $\bar{S}^{tr}[a_{i},B_{\nu}]$ to the associated
renormalized action (\ref{RS.08})
has {\it vanishing} linear in $a_{\mu}$ term,
$\delta\bar{S}^{tr}[a_{i},B_{\nu}]/\delta a_{i}|_{a_{i}=0}=0$ for {\it any}
$B_{\mu}$, that is necessary for self-consistency of the background
perturbation theory. Correspondingly, the
propagator reads $G^{r}_{ij}({\bf y},{\bf x}|B)=g^{2}_{r}{\cal M}^{-2}
<{\bf y}|(\hat{\cal E}^{(0)}_{ij}(B))^{-1}|{\bf x}>$, where
$\hat{\cal E}^{(0)}_{ij}(B)$ is given by eq. (\ref{MR.04}).
As for the difference between the relevant action (\ref{RS.08}) and
${\cal M}^{2}\int d^{D}xTr(a_{i}\hat{\cal E}^{(0)}_{ij}(B)a_{j})/2g^{2}_{r}$,
for a given $L$, it assumes the form of the sum of the counterterms
(\ref{MR.05}) (truncated up to the $L$th order of the expansion) and the
remaining part $\int d^{D}xTr(-(4i[a_{q},a_{l}]D_{q}(B)a_{l}+
[a_{q},a_{l}][a_{q},a_{l}])/4g^{2}_{r}+
\sum_{l=1}^{L}a_{j}{\cal T}_{j}^{(l)}(B))$ of $\bar{S}^{tr}[\cdot]$, with 
the $l\geq{1}$ terms $Tr(a_{j}{\cal T}_{j}^{(l)}(B))$ being treated as
additional vertices.

By virtue of eq.
(\ref{RS.05}), for a given $l$, the role of the latter term is to
exactly cancel, for $\forall B_{\mu}$, the 1PI tad-pole-like subgraphs which
are associated with the $l$th order of the loop-wise expansion of 
$<a_{i}>_{a}^{B}$ evaluated in the theory (\ref{RS.09}). When the condition
(\ref{RS.01}) is violated
already in the classical limit, the self-consistency of the weak-coupling
series (developed for correlators (\ref{GI.06})) is spoiled by the
proliferation of the tree-like subgraphs.
Being generated by the diagrammatic expansion of the
$g^{0}_{r}$th contribution to $<a_{i}>_{a}^{B}$ in the theory (\ref{RS.08}),
they are attached to the rest of a graph by a single 'external'
$a_{\mu}-$line. Proliferation of these
subgraphs is not suppressed: once ${\cal T}_{j}^{(0)}(B)\sim{g^{-2}_{r}}$,
they are {\it all} of the same $g^{0}_{r}$th order.
But, once eq. (\ref{RS.01}) holds true classically, the proliferation of the
associated with $<a_{i}>_{a}^{B}$ subgraphs is already suppressed by powers
of $g^{2}_{r}$.

Finally, the relation (\ref{RS.06}) is consistent with the condition
(\ref{CV.01}). The consistency is maintained provided
$\breve{\Gamma}^{r}[C_{i}|B_{\mu}]=
\breve{\Gamma}^{r}[C^{(\psi)}_{i}|B^{(\psi)}_{\mu}]$ is
invariant under the $a_{i}\rightarrow{C_{i}}$ option of the transformations
(\ref{15.6}) once eq. (\ref{CV.01}) (and, in consequence, eq.
(\ref{15.12v})) is satisfied. In turn, this invariance follows from the
observation that the symmetry (\ref{15.12v}) guarantees the
invariance ${\cal W}^{r}[a^{(\psi)}_{i}|B^{(\psi)}_{\mu}]=
{\cal W}^{r}[a_{i}|B_{\mu}]$ of the associated generating functional
${\cal W}^{r}[J_{i}|B_{\mu}]$ under the transformations (\ref{15.6}).

\section{Renormalizability of the novel representation}
\label{renorm}

Employing the representation (\ref{FM.01m}), we are ready to prove
that the ansatz (\ref{MR.02}), implemented in the tree-order approximation
(\ref{MR.01}), results in the action
$\tilde{S}_{\Lambda_{\varepsilon}}
[a_{i},B_{\mu}|\bar{\vartheta}_{i},{\vartheta}_{i}]$
defining the theory {\it renormalizable} from the power counting
viewpoint. It is most transparent in
the gauge $a_{0}=B_{0}=0$ presumed till the end of the paper.
In this case, for a given normalization point
$\Lambda$, the quadratic in $a_{i}$ and $B_{i}$ part of the tree-level
approximation $\tilde{S}^{tr}_{\Lambda_{\varepsilon}}[\cdot]$ (to the
renormalized action $\tilde{S}^{r}_{\Lambda_{\varepsilon}}[\cdot]$ defined
by eq. (\ref{AC.01})) assumes the form
$\int d^{D}x~Tr(a_{i}\hat{\cal K}^{ij}_{1}a_{j}+
B_{i}\hat{\cal K}^{ij}_{2}B_{j})/2g^{2}_{r}(\Lambda/\Lambda_{YM})$, where
\be
\hat{\cal K}^{ij}_{1}=
\hat{\cal K}^{ij}+
{\cal M}^{2}\delta^{ij}~~~~~,~~~~~
\hat{\cal K}^{ij}_{2}=\hat{\cal K}^{ij}\left(1+
{\triangle}/{{\cal M}^{2}}\right)~~~~,~~~~
\left({\hat{\cal K}^{-1}}\right)^{ij}=\sum_{m=1}^{2}
\left({\hat{\cal K}^{-1}_{m}}\right)^{ij}~,
\label{TR.05}
\ee
where $\hat{\cal K}^{ij}=\hat{P}^{ij}\triangle-
(\delta^{ij}-\hat{P}^{ij})\partial_{0}^{2}$ is the operator defining
(modulo the factor $1/2g^{2}_{r}$) the
quadratic part of the action (\ref{1.1}), while $\hat{P}^{ij}=\delta^{ij}-
\partial^{i}\partial_{l}^{-2}\partial^{j}$, $\triangle=-\partial_{l}^{2}$ and
$(\delta^{ij}-\hat{P}^{ij})\hat{P}^{ij}=0$.
In particular, eq. (\ref{TR.05}) implies that, despite the presence of the
mass-term, among the three components of
the field $a_{i}$ there are only {\it two} propagating polarizations selected
by the projector $\hat{P}^{ij}$.

Next, while the dimensions of the fields are
$[a_{i}]=[\bar{\vartheta}_{i}]=[{\vartheta}_{i}]=1$ and $[B_{i}]=0$,
thus implemented action $\tilde{S}_{\Lambda_{\varepsilon}}[a_{i},B_{i}]$
generates {\it no} vertices with a positive dimension so that
there is only a finite number of correlation functions comprised of
superficially divergent 1PI graphs. Altogether, as it will be sketched in the end
of Section \ref{counterterms}, the pattern of the counterterms reads
\be
\tilde{S}_{CT}[a_{i},B_{i}
|\bar{\vartheta}_{i},\vartheta_{i}]=
\frac{(Z_{g^{2}}^{-1}-1)}{4g^{2}_{r}}\int d^{D}x~Tr
\left(F_{\mu\nu}(a_{i}+B_{i})\right)^{2}~.
\label{MR.05}
\ee
where ${g}_{r}={g}_{r}(\Lambda/\Lambda_{YM})$, and the vector-like ghosts
$\bar{\vartheta}_{i},~\vartheta_{i}$
are treated as independent dynamical fields explicitly
involved in the renormalization algorithm.
It is crucial that the factor $Z_{g^{2}}$, being the same as in
the standard formulation (\ref{1.1}) considered in the gauge $A_{0}=0$,
is accumulated by
the divergent perturbative diagrams {\it without} internal lines associated
either with the low-energy field $B_{i}$ or with the latter ghosts. In
consequence, the counterterms comply with the 
condition
\be
\tilde{S}_{CT}[a_{i},B_{i}|\bar{\vartheta}_{i},\vartheta_{i}]=
\bar{S}^{(pt)}_{CT}[a_{i},B_{i}]~~~~~~~,~~~~~~~~
\tilde{S}_{CT}[a_{i},B_{i}
|\bar{\vartheta}_{i},\vartheta_{i}]=
{S}_{CT}[a_{i}+B_{i}]~,
\label{CT.01v}
\ee
where $\bar{S}^{(pt)}_{CT}[a_{i},B_{i}]$ stands for the counterterms
relevant for the background perturbation theory (i.e., prior to the
integration over $B_{i}$) applied to the averages $<..>_{a}^{B}$ in
the theory (\ref{RS.08}).
The second part of eq. (\ref{CT.01v}) states that
the replacement
$a_{i}+B_{i}\rightarrow{A_{i}}$ transforms the r.h. side of eq.
(\ref{MR.05}) into the well-known pattern of the counterterms
${S}_{CT}[A_{i}]$ evaluated in the framework of the original representation
(\ref{1.1}) in the gauge $A_{0}=0$.

Eq. (\ref{CT.01v}) implies in particular 
that, given the double axial gauge fixing $a_{0}=B_{0}=0$, the
non-renormalization of the involved gauge
fields is valid not only in the background perturbation
theory (for a fixed $B_{i}$) but also in the full
theory (of the two dynamical fields $a_{i}$ and $B_{i}$):
\be
a_{i}^{r}=a_{i}~~~~~,~~~~~B_{i}^{r}=B_{i}~~~~~,~~~~~~
g^{2}=Z_{g^{2}}~{g^{2}_{r}}~.
\label{RS.02}
\ee
The ghost-fields are not renormalized either:
$\bar{\vartheta}_{i}^{r}=
\bar{\vartheta}_{i}$, ${\vartheta}_{i}^{r}={\vartheta}_{i}$. Also,
neither the "bare" mass $M_{a}={\cal M}$ of
$a_{i}$ nor the "bare"
mass $M_{gh}={\cal M}$ of $\bar{\vartheta}_{i},~
\vartheta_{i}$ require any divergent (when
$\varepsilon\rightarrow{+0}$) multiplicative renormalization both prior
and after the intergation over $B_{i}$. In consequence,
the part $S_{m}[a_{i},B_{i}|\bar{\vartheta}_{i},{\vartheta}_{i}]$ of
$\tilde{S}_{\Lambda_{\varepsilon}}
[a_{i},B_{i}|\bar{\vartheta}_{i},{\vartheta}_{i}]$, resulting after the
reformulation (\ref{FM.01m}), contributes to the counterterms neither in
the full theory nor in the background perturbation theory.



Finally, observe that eq. (\ref{TR.05}) displays the basic feature of the
multiscale decomposition: the propagators
$\delta^{ce}{\cal D}^{ij}_{1}({\bf p})$ and
$\delta^{ce}{\cal D}^{ij}_{2}({\bf p})$ of $a^{c}_{i}$ and
$B^{c}_{i}$ ($<{\bf p}|(\hat{\cal K}^{ij}_{k})^{-1}|0>=
{\cal D}^{ij}_{k}({\bf p})$) approach the propagator
$\delta^{ce}{\cal D}^{ij}({\bf p})=\delta^{ce}
<{\bf p}|\hat{\cal K}^{-1}_{ij}|0>$ of the field $A^{c}_{i}$
(of eq. (\ref{1.1})) in the $UV$ and $IR$ domains of the momentum squared respectively.
Owing to the last relation of eq. (\ref{TR.05}), it implies that
${\cal D}^{ij}_{1}({\bf p})>>{\cal D}^{ij}_{2}({\bf p})$ and
${\cal D}^{ij}_{1}({\bf p})<<{\cal D}^{ij}_{2}({\bf p})$ for
${\bf p}^{2}>>{\cal M}^{2}$ and ${\bf p}^{2}<<{\cal M}^{2}$
correspondingly. E.g.,
${\cal D}^{ij}_{1}({\bf p})\sim{({\bf p}^{2})^{0}}$ when
${\bf p}^{2}\rightarrow{0}$, while
${\cal D}^{ij}_{2}({\bf p})\sim({\bf p}^{2})^{-2}$ for
${\bf p}^{2}\equiv p^{2}_{\mu}\rightarrow{\infty}$.

\section{The effective action is of the Wilsonean type}
\label{counterterms}

Given the dimensional regularization $4-D=\varepsilon\rightarrow{+0}$ and
the gauge condition $a_{0}=B_{0}=0$,
let us first adapt the conventional requirement, maintaining that an
effective action is of the Wilsonean type,
to the specific case
(\ref{15.16}) corresponding to
the ansatz (\ref{MR.02}) fixed by eq. (\ref{MR.01}). Secondly, we verify that
the conditions (\ref{CT.01v}) are sufficient to fulfill this requirement.
To begin with, $S_{eff}[B]$ should describe low-energy dynamics separated by
a {\it finite} (for $\varepsilon\rightarrow{+0}$) $UV$ cut off
$\Lambda_{int}$. In our case, $\Lambda_{int}$  is naturally
identified with the $IR$ limit ${\cal M}_{a}={\cal M}/Z^{(pt)}_{{\cal M}}$ of
the renormalized
mass
of the field $a_{i}$ in the auxiliary high-energy theory (\ref{RS.08})
for a fixed $B_{i}$.
Provided ${\cal M}_{a}$ is finite when
$\varepsilon\rightarrow{+0}$ and employing the non-renormalization
(\ref{RS.02}) of $B_{i}=B^{r}_{i}$,
the requirement reads: the operator expansion of this action is expressed,
$S_{eff}[B]=\sum_{n\geq{1}}c_{n}({\cal M}_{a})
~{\cal O}_{n}[B]$, in terms of $B_{i}$ and 
{\it renormalized} coupling constants $c_{n}({\cal M}_{a})$. The
coefficients $c_{n}({\cal M}_{a})$ are
given by the $\bar\Lambda={\cal M}_{a}$ option of the
"running" constants $c_{n}(\bar\Lambda)$ which, for any
$\varepsilon-$independent $\bar\Lambda$, should possess a finite limit
when $\varepsilon\rightarrow{+0}$.
It means that the effective theory is free of $UV$
divergences which are regularized due to an implicit $UV$ cutoff of order of
${\cal M}_{a}$ implemented by the action $S_{eff}[B]$.


In the auxiliary theories (\ref{RS.08}) and (\ref{RS.09}) considered for a
fixed $B_{i}$, the $IR$ limit 
${\cal M}_{a}\equiv{\cal M}_{a}({\cal M},\Lambda,\Lambda_{YM})$
of the renormalized $a_{i}-$mass is defined (see below) by the
relation
\be
{{\cal M}_{a}}/
{\tilde{g}_{r}({\cal M}_{a}/\Lambda_{YM})}=
{{\cal M}}/{{g}_{r}(\Lambda/\Lambda_{YM})}~,
\label{CT.01k}
\ee
where $\tilde{g}^{2}_{r}({\cal M}_{a}/\Lambda_{YM})$ denotes the $IR$ limit
(to be introduced after eq. (\ref{EA.32q})) of the coupling constant
in the latter auxiliary theories. Therefore, the
scale ${\cal M}_{a}=\Lambda_{int}$ of the interpolation is {\it finite}
in the limit $\varepsilon\rightarrow{+0}$ provided both ${\cal M}$ and
$\Lambda$ are chosen to be $\varepsilon-$independent (in compliance with
the conditions (\ref{CT.01v})).
Once ${\cal M}_{a}$ is finite, the above requirement on $S_{eff}[B]$ is
tantamount to the first of the conditions
(\ref{CT.01v}) imposed on the counterterms of the {\it microscopic}
theory determined by the
(conventionally renormalizable)
action $\tilde{S}_{\Lambda_{\varepsilon}}[a_{i},B_{i}]$.
Indeed, it justifies that in the
$\varepsilon\rightarrow{+0}$ limit the
effective theory is free of $UV$ divergences. Correspondingly, the action
$S_{eff}[B]=\sum_{n\geq{1}}c_{n}({\cal M}_{a}) ~{\cal O}_{n}[B]$ is 
$1/\varepsilon-$independent when expressed in terms of
${\cal M}_{a}$, $B_{i}=B^{r}_{i}$ and $g^{2}_{r}=g^{2}/Z_{g^{2}}$, where
$Z_{g^{2}}$ is defined by eq. (\ref{MR.05}).

Applying the renormalized background perturbation theory
(combined with the covariant derivatives' expansion),
the computation of the effective action (\ref{15.16}) considerably
simplifies when, in the theory (\ref{RS.08}), the $IR$ limit
${\cal M}_{a}={\cal M}_{a}({\cal M},\Lambda,\Lambda_{YM})$ of the
$a_{i}-$mass coincides with the tree-level approximation ${\cal M}$ to this
mass. To this aim, one is to select such
$\Lambda=\breve{\cal M}$ that
\be
{g}^{2}_{r}(\breve{\cal M}/\Lambda_{YM})=
\tilde{g}^{2}_{r}({\cal M}/\Lambda_{YM})~~~~~
\Longleftrightarrow~~~~~
{\cal M}_{a}({\cal M},\breve{\cal M},\Lambda_{YM})={\cal M}~.
\label{CT.01kk}
\ee
Introducing the reparameterization $\Lambda\rightarrow
\tilde{\Lambda}(\Lambda)\equiv
\tilde{\Lambda}(\Lambda,\Lambda_{YM})$
via the relation ${g}^{2}_{r}(\Lambda/\Lambda_{YM})=
\tilde{g}^{2}_{r}(\tilde{\Lambda}/\Lambda_{YM})$, we obtain
$\breve{\cal M}=\tilde{\Lambda}^{-1}({\cal M})$.
In accordance with the concept of the anomalous dimension, the relation
(\ref{CT.01k}) between ${\cal M}_{a}$ and ${\cal M}$ implies then that, for
a fixed ${\cal M}_{a}$ and $\forall\Lambda\geq\breve{\cal M}=
\tilde{\Lambda}^{-1}({\cal M}_{a})$,
the quantity ${\cal M}={\cal M}(\Lambda)$ can be reinterpreted as the running
mass associated with the scale
$\Lambda$ (with ${\cal M}(\breve{\cal M})={\cal M}_{a}$).



In conclusion, let us sketch the derivation of 
eqs. (\ref{CT.01v}) and (\ref{CT.01k}).
To justify the second of the conditions (\ref{CT.01v}),
it is convenient to treat the action
$\tilde{S}_{\Lambda_{\varepsilon}}
[a_{i},B_{\mu}|\bar{\vartheta}_{i},{\vartheta}_{i}]$,
defined by the trick (\ref{RE.05b})/(\ref{MR.02}) together with
the specification (\ref{MR.01}) of ${\cal T}_{j}(B)$,
as belonging to the {\it two}-parametric variety. For this purpose,
$S^{r}_{m}[\cdot]$ is generalized to $S^{r}_{m}[a_{i},B_{\mu}|\{\xi_{k}\}]$
so that, in eq. (\ref{RE.08}), $w_{\mu}$ is
replaced by $\xi_{1}a_{\mu}+
\xi_{2}g^{2}_{r}{\cal T}_{\mu}(\cdot)/{\cal M}^{2}$.
As the considered implementation of the transformation (\ref{15.14}) keeps
intact the renormalizability in the power counting sense, there should exist
such multiplicative renormalization both of $\xi_{k}=Z_{\xi_{k}}\xi^{r}_{k}$
and of $\bar{\vartheta}_{i}=Z_{\bar{\vartheta}}\bar{\vartheta}_{i}^{r}$,
${\vartheta}_{i}=Z_{{\vartheta}}{\vartheta}_{i}^{r}$ that,
together with eq. (\ref{RS.02}), allows to separate the
relevant counterterms
$\tilde{S}_{CT}[a_{i},B_{\mu}|\bar{\vartheta}_{i},\vartheta_{i}]$
to cancel all the $UV$ divergences in the theory with thus specified 
$\tilde{S}_{\Lambda_{\varepsilon}}[\cdot|\cdot]$. In view of the
non-renormalization (\ref{RS.02}) of the fields (which holds true
by virtue of the non-renormalization
$A_{i}=A^{r}_{i}$ in the axial gauge option of the original formulation
(\ref{1.1})) the difference between the ghost-independent counterterms
$\tilde{S}_{CT}[a_{i},B_{i}|0,0]$ and
${S}_{CT}[a_{i}+B_{i}]$ may be composed (by virtue of eq. (\ref{AC.01})
only of 
$\beta_{1}=\int d^{D}x~Tr(a^{2}_{i})$,
$\beta_{2}=\int d^{D}x~Tr(a_{i}~\delta S_{YM}[B_{k}]/\delta B_{i})$, and
$\beta_{3}=\int d^{D}x~Tr((\delta S_{YM}[B_{k}]/\delta B_{i})^{2})$.
In the two-parametric variety,
to guarantee that $\tilde{S}_{CT}[a_{i},B_{i}|0,0]=
{S}_{CT}[a_{i}+B_{i}]$, it is sufficient to verify the absence of the
counterterms proportional to any two $\beta_{q}$. We choose $q=1,3$
which, in
particular, would imply that
$\delta M^{2}_{a}=0$, where $\delta M^{2}_{a}$ denotes the divergent 
renormalization of the squared mass $M^{2}_{a}$ of $a_{i}$.
On the other hand, the power-counting proves that a possible
ghost-dependent part of
$\tilde{S}_{CT}[a_{i},B_{i}|\bar{\vartheta}_{i},\vartheta_{i}]$
may be associated only with the renormalization
$\delta M^{2}_{gh}$ of the mass ${\cal M}$ of the ghosts which is excluded 
since the ansatz (\ref{MR.02}) guarantees that
$\delta M^{2}_{gh}=\delta M^{2}_{a}=0$.

Concerning the required verification, the power-counting demonstrates that,
in the considered variety of the theories, the divergent
perturbative diagrams do not generate the combination $\beta_{3}$. To justify
that the remaining combination
$\beta_{1}\equiv\beta_{1}[a]$
is not generated either, we make the inverse change
${\cal D}B_{i}{\cal D}a_{i}\rightarrow{{\cal D}B_{i}{\cal D}A_{i}}$ of the
variables to show that the difference
$\Delta\tilde{S}_{CT}[A_{i},B_{i}]=
\tilde{S}_{CT}[A_{i}-B_{i},B_{i}|0,0]-{S}_{CT}[A_{i}]$ may be only such
functional that $\Delta\tilde{S}_{CT}[A_{i},0]=0$ for $\forall{A_{i}}$
(which excludes $\beta_{1}[A-B]$ since $\beta_{1}[A-B]|_{B=0}\neq{0}$ for
$A^{2}_{i}\neq{0}$). For this purpose, consider the generating functional
${\cal W}[J^{+}_{i},I_{i}]$ (with ${\cal W}[0,0]=0$)
which results after the averaging of
$e^{\int d^{D}x~Tr(J^{+}_{i}A_{i}+I_{i}B_{i})}$
in the theory defined by the action
$\tilde{S}_{\Lambda_{\varepsilon}}[A_{i}-B_{i},B_{i}]$
(implicitly depending on $\{\xi_{k}\}$). Integration over $B_{i}$ yields
$e^{{\cal W}[J^{+}_{i},I_{i}]}=<e^{\int d^{D}x~Tr(J^{+}_{i}A_{i})}>_{A}^{I}$
where the $A_{i}-$averaging is performed with respect to the action
$({S}_{YM}[A_{i}]-\Delta\tilde{\cal W}[I_{i}|A_{i}])$, and
$e^{\Delta\tilde{\cal W}[I_{i}|A_{i}]}=\int {\cal D}B_{i}~
e^{-{S}_{m}[A_{i}-B_{i},B_{i}|\{\xi_{k}\}]+
\int d^{D}x~Tr(I_{i}B_{i})}$. Eq. (\ref{RE.05b})
leads to $\Delta\tilde{\cal W}[0|A_{i}]=0$ which, in turn,
implies the required condition
$\Delta\tilde{S}_{CT}[A_{i},0]=0$. Indeed, the constraint
$\Delta\tilde{\cal W}[0|A_{i}]$ means that 
$\Delta\tilde{S}_{CT}[A_{i},B_{i}]$ may be generated only by those 1PI
diagrams which necessarily possess a {\it nonzero} number of external
$B_{i}-$lines. These lines are associated with subgraphs 
composed into correlation functions which are
obtained applying
the functional $\delta /\delta I_{i}({\bf x})-$derivatives to the
intermediate generating functional
$\Delta\tilde{\cal W}[I_{i}|A_{i}]$.


To justify the first of the conditions (\ref{CT.01v}), we prove
that $\bar{S}^{(pt)}_{CT}[a_{i},B_{i}]={S}_{CT}[a_{i}+B_{i}]$ which,
in view of the identity\footnote{This identity follows from eq. (\ref{RS.06})
and the relation $\Gamma^{r}[C_{i}|B_{\mu}]=\breve\Gamma^{r}[C_{i}|B_{\mu}]+
\int d^{D}x~Tr(C_{j}{\cal T}_{j}(B))$ introduced prior to eq. (\ref{RS.06}).}
$\breve{S}^{(pt)}_{CT}[a_{i},B_{i}]=\bar{S}^{(pt)}_{CT}[a_{i},B_{i}]$,
is a consequence
the condition $\breve{S}^{(pt)}_{CT}[a_{i},B_{i}]={S}_{CT}[a_{i}+B_{i}]$,
where $\breve{S}^{(pt)}_{CT}[\cdot]$ denotes the counterterms in the
theory defined, for a fixed $B_{i}$, by the action
(\ref{RS.09}). To verify the latter condition,
let us temporarily omit the contribution associated with
the mass term of the $a_{i}-$field. Then, it is easy to derive
that the corresponding Legendre effective action (LEA)
$\breve{\Gamma}[C_{i}|B_{i}]|_{{\cal M}=0}=
{\Gamma}_{YM}[C_{i}+B_{i}]$, where 
${\Gamma}_{YM}[C_{i}]$ denotes LEA in the
theory (\ref{1.1}) in the gauge $A_{0}=0$.
In consequence, $\breve{S}^{(pt)}_{CT}[a_{i},B_{i}]|_{{\cal M}=0}=
{S}_{CT}[a_{i}+B_{i}]$, and eq. (\ref{RS.02}) remains valid.
Reintroducing the mass term into eq. (\ref{RS.09}), the power counting shows
that
the difference $\breve{S}^{(pt)}_{CT}[a_{i},B_{i}]-
\breve{S}^{(pt)}_{CT}[a_{i},B_{i}]|_{{\cal M}=0}$ may be associated only with
a possible renormalization $\delta^{(pt)} M^{2}_{a}$ of the mass of the field
$a_{i}$. In turn, $\delta^{(pt)} M^{2}_{a}=0$ by virtue of eq. (\ref{CT.01k}).

To prove eq. (\ref{CT.01k}),
consider the generating
functional $\breve{\cal W}_{1}[J_{i}|{\cal V}_{i}]$ which
results after the averaging of the source
$e^{\int d^{D}x~Tr(J_{i}a_{i})}$ in the auxiliary high-energy theory (of the
single dynamical field $a_{i}$) defined by the
action $\breve{S}[a_{i}-{\cal V}_{i},{\cal V}_{i}]$ depending
on the external field ${\cal V}_{i}$ so that the normalization is chosen to be
$\breve{\cal W}_{1}[0|0]=0$. Similarly to \cite{Abb}, one justifies
that the ${\cal V}_{i}=C_{i}$ option of the associated LEA
$\breve{\Gamma}[C_{i}|{\cal V}_{i}]$ (with $C_{i}$ being
conjugated to $J_{i}$) is the $C_{0}=0$ reduction of a gauge-invariant
functional: $\breve{\Gamma}[C_{i}|C_{i}]=\hat{\Gamma}[C_{i}]$ with
$\hat{\Gamma}[F_{\mu}]=\hat{\Gamma}[F^{(\omega)}_{\mu}]$.
Owing to the latter property of $\breve{\Gamma}[C_{i}|C_{i}]$,
it is the leading term $\int d^{D}x~tr(F^{2}_{\mu\nu}(C_{i}))/4
\tilde{g}^{2}_{r}({\cal M}_{a}/\Lambda_{YM})$ of the operator expansion of
$\breve{\Gamma}[C_{i}|C_{i}]$ that defines
the coupling constant  $\tilde{g}^{2}_{r}({\cal M}_{a}/\Lambda_{YM})$. In turn,
to evaluate ${\cal M}_{a}$, one notes that the leading term of the operator
expansion of $\breve{\Gamma}^{r}[C_{i}|0]$ assumes the form
${\cal M}^{2}_{a}\int d^{D}x~tr(C^{2}_{i})/2
\tilde{g}^{2}_{r}({\cal M}_{a}/\Lambda_{YM})$. Altogether, the
condition (\ref{CT.01k}) follows (in view of $C_{i}=C^{r}_{i}$) from the
${\cal V}_{i}=C_{i}$ option of the
relation
\be
\breve{\Gamma}^{r}[C_{i}|{\cal V}_{i}]=
\breve{\Gamma}^{r}[C_{i}|0]+{{\cal M}^{2}}\int d^{D}x~
Tr\left({\cal V}^{2}_{i}-2C_{i}{\cal V}_{i}\right)
/{2g^{2}_{r}(\Lambda/\Lambda_{YM})}~,
\label{EA.32q}
\ee
following from the fact that,
in the action 
$\breve{S}[a_{i}-{\cal V}_{i},{\cal V}_{i}]$,
the coupling between ${\cal V}_{i}$ and $a_{i}$ is linear in
$a_{i}=a^{r}_{i}$. Indeed, in the theory defined by
$\breve{S}[a_{i}-{\cal V}_{i},{\cal V}_{i}]$, consider
the renormalized perturbative expansion
in ${g}_{r}(\tilde{M}/\Lambda_{YM})$
with some $\tilde{M}>\Lambda_{YM}$ so
that, at the tree-level, the $a_{i}-$mass is equal to
$\tilde{\cal M}_{a}={\cal M}\tilde{g}_{r}(\tilde{\Lambda}(\tilde{M})/
\Lambda_{YM})/{g}_{r}(\Lambda/\Lambda_{YM})$, where $\tilde{\Lambda}(\Lambda)$ is defined after eq. (\ref{CT.01kk}). Given the definition
(\ref{CT.01k}) of ${\cal M}_{a}$,
it is the choice $\tilde{M}=\tilde{\Lambda}^{-1}({\cal M}_{a})$ (resulting in
$\tilde{\cal M}_{a}={\cal M}_{a}$)
that, in view of eq.
(\ref{EA.32q}), allows to fix
the $IR$ {\it limit} both of the $a_{i}-$mass and of $\tilde{g}^{2}_{r}$
already in the tree-approximation to the action (\ref{RS.09}).
In this case, the $n\geq{1}$ loop contributions to the
coefficient of the leading term of the operator expansion of
$\breve{\Gamma}[C_{i}|C_{i}]$ are exactly
cancelled by the counterterms (\ref{MR.05}).


\section{Conclusions}

Building on the insertion of the unity (\ref{15.13}), we propose
the multi-scale decomposition (\ref{15.14}) which respects
the background gauge invariance (\ref{15.6}) and
resolves, via eq. (\ref{RS.06}), the constraint (\ref{RS.01}) up to any given
order of the loop-wise expansion. Choosing the axial gauge (\ref{15.14i}) and
employing the ansatz (\ref{MR.02})/(\ref{MR.01}),
it introduces a novel renormalizable representation of the gauge theory
(\ref{1.1}). In turn, it allows to synthesize qualitatively
different methods to evaluate the contribution of the high- and low-energy
fields $a_{\mu}$ and $B_{\mu}$ interpolated at a scale $\Lambda_{int}$. The
first average $<..>_{a}^{B}$ is performed employing the
$1/N$ weak-coupling expansion associated with a normalization point $\Lambda$.
Due to the presence of the $a_{i}-$mass term (\ref{RS.09}) at the tree-level,
it does not exhibit spurious $IR$ singularities (present
when this expansion is applied directly to the original formulation
(\ref{1.1})) provided the $IR$ limit
${\cal M}_{a}({\cal M},\Lambda,\Lambda_{YM})=\Lambda_{int}$ of the
renormalized mass, defined by eq. (\ref{CT.01k}), is sufficiently larger than
$\Lambda_{YM}$.
Integrating over $a_{\mu}$, an arbitrary correlator
${\cal Q}[A_{\mu}]={\cal Q}[A^{(\omega)}_{\mu}]$ is expressed, by virtue of
eq. (\ref{15.12v}), in
terms of gauge-invariant generically non-local correlators (i.e., Wilson loops with various
operator's insertions) averaged
with the $B_{\mu}-$dependent effective action (\ref{15.16}). Respecting the
gauge symmetry,
the latter action is verified to be of the Wilsonean type. We also note that
application of analytical approximations to the computation of
$S_{eff}[B]$ is considerably
facilitated by the judicious adjustment (\ref{CT.01kk}) between the
parameters ${\cal M}_{a}$ and $\Lambda$ when the proposed Ansatz depends
(in addition to $\Lambda_{YM}$) on the
{\it single} parameter ${\cal M}_{a}={\cal M}$.


As the low-energy theory is supposed to be strongly coupled, one possible way
to evaluate the low-energy correlators is to
develop further the stringy representation of the $1/N$ strong-coupling
expansion introduced in \cite{Dub3} (see also \cite{Dub&Mak}) for the
continuous $D=4$ Yang-Mills theory. The interpolation between the $1/N$
strong- and $1/N$ weak-coupling series suggests that the gauge theory
can be represented in a synthetic way\footnote{It can be compared to
the output of the semi-phenomenological approach \cite{BPT} (corresponding to
an implementation of the trick (\ref{15.13}) with some unspecified
$a_{\mu}-$independent action $\tilde{S}_{m}[a_{\mu},B_{\mu}]=
\tilde{S}_{m}[B_{\mu}]$) where it is argued that massless, at least
at the tree-level, gluons are coupled to a "frozen" string.}
combining "massive" gluons (with {\it two} propagating components) and
fluctuating confining strings. Indeed, let identify
$e^{{\cal R}[A_{\mu}]}$ in eq.
(\ref{15.14}) with a macroscopic Wilson loop $W_{C}$. Then, appropriate
segments of the base-contour $C$ are collected, together with the trajectories
of the $a_{\mu}-$gluons, into {\it closed} auxiliary contours ${\cal C}_{k}$
which constitute boundaries of strings associated with the
representation of the correlators like
$<\prod_{k}W_{{\cal C}_{k}}>_{B}$. A work in this direction is in progress.


\enddocument